\documentclass{aa} 
\usepackage{natbib}
\usepackage{amssymb}

\newcommand{\ds}{\displaystyle}
\newcommand{\inv} {\frac {1}}

\newcommand{\derivp} [2] {\frac {\partial #1 } {\partial #2} }
\newcommand{\deriv} [2] {\frac {d #1 } {d #2} }
\newcommand{\dprime}  {{\prime \prime}}
\newcommand{\eqn} [1] {
\begin{equation}#1
\end{equation}}
\newcommand{\eqna} [1] {
\begin{eqnarray}#1
\end{eqnarray}}
\newcommand{\ie} {{\sl i.e. }}

\begin{document}


\title{Excitation of stellar p-modes  by turbulent convection:}

\subtitle {1. Theoretical formulation }

\author{Samadi R. \inst{1} \and   Goupil M.-J.\inst{2}  }

\institute{
Observatoire de Paris. DESPA. CNRS UMR 8632. 92195 Meudon. France \and
Observatoire de Paris. DASGAL. CNRS UMR 8633. 92195 Meudon. France
}
\offprints{R. Samadi, \email{Reza.Samadi@obspm.fr}}

\date{Received 17 October 2000/ Accepted 19 January 2001}

\abstract{Stochatic excitation of stellar oscillations by  turbulent 
convection is investigated and an expression for the power injected 
into the oscillations by
the turbulent convection of the outer layers is derived which takes into
account excitation through turbulent Reynolds stresses 
and turbulent entropy fluctuations.  
This formulation generalizes results from previous works and 
is built so as to enable investigations of  various
 possible spatial and temporal 
spectra of stellar turbulent convection.
For the Reynolds stress contribution and assuming  
the Kolmogorov spectrum we obtain a similar formulation 
to those derived by previous authors.
The entropy contribution to excitation is found to originate from the advection
of the Eulerian entropy fluctuations by the turbulent velocity field. Numerical
computations in the solar case in a companion paper  
indicate that the entropy source term
is dominant over the Reynold stress contribution to mode excitation, 
except at high frequencies. 
}
\maketitle

\markboth{}{}

\section{Introduction}

Oscillation amplitudes and linewidths provide
information on  excitation and damping processes of stellar
oscillation modes. 
In the solar case, the observed oscillation modes are believed to be 
 damped as a result of a competition between several non-adiabatic and turbulent
processes \citep{Osaki90,Houdek99}. Excitation of solar oscillation modes is 
attributed to turbulent convection at the surface of the Sun.
The first theoretical investigation of solar oscillation excitation 
by turbulent convection was by \citet[hereafter GK]{GK77}. 
These authors identified the turbulent term 
of Reynolds stress as the main source term of stochastic  excitation of 
solar acoustic modes in the wave equation.  
GK derived an approximate estimation for
the acoustic power injected into the oscillations by  turbulent 
convection which arises from an equipartition of energy between the 
turbulent elements and the oscillations. 
The result, however, underestimated the power
 by a factor  $\sim 10^3$ compared to the solar observations
 \citep{Osaki90}. 
Amplitude estimations for solar-like oscillating stars have subsequently 
been  computed by   \citet{JCD83a} 
based on this simple picture of 
equipartition of energy between turbulence and oscillation.
GK considered the adiabatic assumption for both the 
oscillations and the turbulence.
 A decade later, entropy 
  turbulent  fluctuation 
 has been proposed as a possible  additional excitation source 
 \citep{Stein91,Balmforth92c,GMK94}. 

The stochastic mechanism may  be understood as follows~:
turbulent motions of the stellar material and turbulent fluctuations 
of thermodynamicals quantities, which occur in the convection zone, 
generate acoustic waves \citep{Lighthill52,Stein67}.
The corresponding acoustic power then excites resonant modes of the 
stellar cavity (oscillations). This excitation of the oscillations  
results from  a forcing by  incoherent (turbulent) source terms due 
to the turbulent Reynolds stress and  turbulent entropy fluctuations.

An alternative  formulation for the power and amplitude 
oscillation is proposed by   \citet[hereafter B92]{Balmforth92c} and is used  by \citet{Houdek99} to compute 
amplitude of oscillation for various  solar-type stars.
 There is some disagreement between the conclusions of  both formulations concerning
the entropy contribution~: 
it is found
 dominant  in theoretical investigation 
 \cite[hereafter GMK]{GMK94} and as a result 
of numerical 3D simulations \citep{Stein91}, 
but  negligible in some other cases  \citep{Balmforth92c}.

Both formulations  are
built following the  method developed by  GK 
and are based on a  simplified description of the turbulent medium. 
The spectrum of turbulent energy in stellar conditions 
is, however, not well known. 
The best-known spectrum   is, of course, the solar spectrum 
\citep{Espagnet93,Nesis93} and observations tell us that a Kolmogorov spectrum does 
not represent the complete turbulent solar spectrum \citep{Nesis93}.

The purpose of the present work  therefore is
  to investigate the effect  of several 
possible models of   turbulence on the excitation
of solar-like oscillation modes and  
to establish their  signature in power spectra. 
To do so, a formulation must first be established which
includes both (kinetic and entropic) contributions in a consistent
  and general way.
The main goal is to allow the use of  any form of 
the turbulent kinetic energy spectrum, 
of turbulent entropy fluctuations
and the eddy temporal spectrum.
Two free parameters are introduced to take 
into account uncertainties in the definitions of 
the coherence time of the turbulent eddies and  in the magnitude
of the wavenumber at which the inertial regime starts.

Once the general formulation is established, we study the specific case of a
  Gaussian time  and Kolmogorov energy spectra, which enables us to compare our findings with results of previous works.  In a companion paper, 
the present formulation  is  applied to the solar case
and several turbulent models found in the
  literature are discussed. Both free parameters can be
 calibrated with solar observations 
and used to compute oscillation amplitudes for  other
 potential solar-like oscillating   stars
\citep{Samadi00b,Samadi00III} in view of forthcoming space seismological
  experiments~: COROT  \citep{Baglin98}, MONS \citep{Kjeldsen98}, MOST \citep{Matthews98}, EDDINGTON \citep{Favata00}.

Sect.~1 recalls  how the 
stochastic excitation mechanism of stellar oscillations is modeled by 
source terms   in the wave equation.
We assume the modes are damped and  find that they  are excited through
 the turbulent Reynolds stresses \citep{GK77} and  the turbulent 
entropy fluctuations  arising from the turbulent nature of 
the stellar convection region \citep{Stein91,Balmforth92c,GMK94}. 
We adopt the GK procedure and  assume
 that the oscillation modes are decoupled from the turbulent medium.
This description gives rise to an  inhomogeneous 
wave  equation 
for the oscillation modes. In this framework, the acoustic turbulent source
acts as a forcing term for the oscillation normal modes. 
We consider  adiabatic radial oscillations in 
the \citet{Cowling41} approximation. 
A  homogeneous, isotropic, stationary  turbulence
 is assumed. A  formulation for the Reynolds stress contribution 
which can include any type of  turbulent spectrum 
is established in Sect.~3.   
In Sect.~4,  the  contribution  of turbulent 
fluctuations  of entropy is worked out.
 We find that it is the  advection of the
entropy turbulent fluctuation by the turbulent velocity 
field  which yields the entropy 
source.
An appendix shows that cross terms 
between Reynolds and entropic sources do not contribute.

Sect.~5 finally establishes a general  formulation 
which can be used to investigate any type  of turbulent  spectrum. 
Finally, Sect.~6 discusses the complete  formulation, its limitations and 
advantages.

\section{Turbulent stochastic excitation}

\subsection{The inhomogeneous wave equation}

In the basic hydrodynamical equations, we use the symbols $P$, $\rho$,
 $\vec v$ and $\vec g$ to denote respectively pressure, 
density, velocity and gravitational acceleration. 
Equilibrium quantities are represented with a subscript 0.
 Each variable $f$, except for the velocity $\vec v$,  is written   as 
the sum of the equilibrium quantity , $f_0$ and a Eulerian  fluctuation, $f_1$,
$f = f_0 + f_1$ and 
we retain terms which are linear and quadratic in the variables 
$P_1$ and $\rho_1$ and  neglect the gravitational perturbation.
Accordingly, one  obtains for the perturbed momentum
and continuity equations:
\eqna{
\derivp{\rho \vec v  } {t}  + \vec \nabla : ( \rho \vec v \vec v ) + 
\vec \nabla P_1 - \rho_ 1 \vec g_0   = 0
\label{eqn:perturbed_momentum_eqn}
\\
\derivp{ \rho_1} {t} + \vec \nabla . (  \rho \vec v) = 0 \; .
\label{eqn:perturbed_continuity_eqn}
}

The perturbed  equation of state in a Eulerian description is given by:
\eqna{
P_1 = c_s^2 \rho_1 +    \alpha_s  s_1 + \alpha_{\rho \rho} \rho_1 ^2
 + \alpha_{ss}  s_1^2 + \alpha_{\rho s} \rho_1 s_1
\label{eqn:perturbed_state_eqn}
}
where  $s$ is the entropy, $\displaystyle{\alpha_s =\left ( \partial P /\partial s  \right )_\rho}$,  $\ds{c_s = \Gamma_1 \,  P_0/\rho_0}$  denotes the  average sound speed, $\displaystyle{\Gamma_1 = \left( \partial \ln P / \partial \ln \rho \right )_s }$ is the  adiabatic exponent and $\alpha_{\rho \rho} $, $\alpha_{ss}$ and $\alpha_{\rho s}$ are the second partial derivatives of $P$ versus $s$ and $\rho$.

We assume  adiabatic oscillations~: the 
Lagrangian entropy fluctuations are therefore only due to turbulence. 
Lagrangian and Eulerian  
entropy fluctuations are related to each other
 by
\eqna{
\deriv{\delta s_t}{t}  = \derivp {s_1} {t}  +  \vec v . \vec \nabla 
(s_0 +  s_1)
\label{eqn:lagran_euler_s}
}
where $\delta s_t $ is the turbulent Lagrangian entropy fluctuation.
The subscript $t$ refers  to  turbulent quantities.
For later use and with the help of Eq.(\ref{eqn:lagran_euler_s}), 
the time derivative of the Eulerian entropy fluctuation $s_1$ is written as
\eqna{
\alpha_s \derivp{s_1}{t} = \alpha_s \deriv{\delta s_t}{t}  - 
\alpha_s \, \vec v . \vec \nabla s_0 - \vec \nabla . ( \alpha_s s_1 \vec v ) 
  \nonumber \\ 
+ s_1 \vec v . \vec \nabla \alpha_s + \alpha_s s_1 \vec \nabla . \vec v \; .
\label{eqn:derivp_s1}
}
The velocity field $\vec{v} $ is split into a component due to the  
pulsational displacement $\delta \vec r_{osc} $  and a 
 turbulent  component $\vec{u}$ as $\vec{v} =\vec v_{osc}  +  \vec{u}$.

 Linearisation of Eq.(\ref{eqn:perturbed_momentum_eqn}-\ref{eqn:perturbed_state_eqn}) yields for the velocity field, 
in the absence of turbulence ($\vec u = 0$), the homogeneous wave equation
\eqna{
\left ( \derivp  { ^2 } {t^2}  -  \vec L  \right )  
\vec {v}   = 0
\label{eqn:homogeneous_wave}
}
with the linear wave  operator~:
\eqna{
\vec L(\vec X)  =  \inv{\rho_0} \left [ \vec \nabla \left ( c_s^2 
\vec \nabla . ( \rho_0 \vec X) + \alpha_s \vec X . \vec \nabla s_0 \right )
\right .  \nonumber \\
 \left . - \vec g_0  \vec \nabla . ( \rho_0 \vec X) \right ] \; .
}
With  appropriate boundary conditions \citep{Unno89}
one recovers 
 the usual eigenvalue problem~:
\eqna{
\vec L(\vec \xi(\vec r))  =  - ~ \omega_0^2 \; \vec \xi(\vec r)
}
where  $\omega_0$ is the oscillation frequency and $\vec \xi(\vec r)$  is the
adiabatic 
(real) displacement  eigenvector.

In the presence of turbulence, the pulsational displacement  and  velocity are 
 written   in terms of the above adiabatic 
solution $\vec {\xi} (\vec{r},t)$  and an instantaneous
amplitude   $A(t)$.
Accordingly 
\eqn{
\vec v_{osc} = \deriv{\delta \vec r_{osc}} {t} 
=  {1\over 2} 
(-i \omega_0 \,  A(t) \,  \vec {\xi} (\vec r) \,  e^{-i \omega_0 t } +cc)
\label{eqn:vosc}
}
where cc means complex conjugate.

Differentiating Eq. (\ref{eqn:perturbed_momentum_eqn}) 
with respect to $t$, neglecting  non-linear terms in $\vec v_{osc}$,
 assuming an incompressible turbulence ($\vec \nabla . \vec u =0$)   
and using
Eq. (\ref{eqn:perturbed_continuity_eqn}, \ref{eqn:perturbed_state_eqn}, \ref{eqn:derivp_s1})
 yields  the inhomogeneous wave equation 
\eqna{
\rho_0 \left ( \derivp  { ^2 } {t^2}  -  \vec L  \right )
\left [ (1+\frac{\rho_1}{\rho_0}) \vec v  \right ] +
\vec {\cal D} (\vec v_{osc})  & = & \nonumber  \\
& & \hspace{-1.5cm}
 \derivp{}{t}
\left (  \vec f_{t}  + \vec \nabla  h_t   \right ) 
\label{eqn:inhomogeneous_wave}
}
with
\eqn{
 \vec f_{t} = -\vec \nabla :  (\rho_0\vec {u} \vec{u} )
\label{eqn:f_t}
}
the turbulent Reynolds stress,
\eqna{
\derivp{}{t} \vec \nabla h_t 
=  -  \vec \nabla \left  ( \right . \alpha_s \deriv{}{t} \delta s_t - 
\vec \nabla. (\alpha_s \vec u \, s_1 ) +
 s_1 \vec u . \vec \nabla \alpha_s \left . \right ) \; ,
\label{eqn:h_t}
}
the source term due to the turbulent entropy fluctuations.
The first term in the RHS of Eq. (\ref{eqn:h_t}) is due to the Lagrangian
entropy fluctuation which  was considered by GMK and B92. 
The last two  terms are due to the buoyancy force associated with the Eulerian 
entropy fluctuations. These  terms contribute  to the excitation
as much as  the Reynolds source term.
The time derivative operator in the LHS of Eq. (\ref{eqn:h_t}) in the
definition of $h_t$ is introduced 
for convenience.

The operator $\vec {\cal D}$ involves both the turbulent velocity field
($\vec u$) 
and the pulsational velocity and is defined as 
\begin{eqnarray}
 \vec {\cal D}  (\vec v_{osc}) &
 =  & 2 \derivp{}{t} \left (\vec \nabla : \rho \vec u \, \vec v_{osc} \right )  +
\vec \nabla \Bigl ( \Bigr . \alpha_s \frac{\rho_1}{\rho_0} \vec v_{osc} . 
\vec \nabla s_0
\nonumber 
\\ 
& & - \vec \nabla . ( \alpha_s \vec v_{osc} \, s_1)  
 +  s_1 \vec v_{osc} . \vec \nabla \alpha_s 
      \\ 
& &  +  \alpha_s s_1 \vec \nabla . \vec v_{osc} 
 -  (\alpha_{\rho s} s_1  +  2 \alpha_{\rho \rho} \rho_1) 
\vec \nabla . ( \rho \vec v_{osc} )  
 \Bigl . \Bigr ) \nonumber .
\label{eqn:D}
\end{eqnarray}
As  will be shown in the next section, this term contributes to the 
dynamical linear damping.

We assume a ``free turbulence'', {\sl i.e.} the turbulent medium evolves 
freely and is not perturbed by the oscillations. 
The continuity equation is  verified by the turbulent medium  such that
\eqna{
 \derivp{ \rho_t}{t} + \vec \nabla .  
\left (  ( \rho_0 + \rho_t ) \, \vec u \right ) = 0 \; .
\label{eqn:continuity}
}
As a consequence of  neglecting the oscillation perturbations in higher order terms, 
the Eulerian fluctuations $s_1$ and $\rho_1$ can be seen as due only to the
turbulence in evaluating the quantities 
 $\vec \nabla  h_t$ and $\vec{\cal D}$. 
We therefore  substitute  $\rho_1$ (resp. $s_1$) by $\rho_t$ (resp. $s_t$) 
 in Eq. (\ref{eqn:h_t},\ref{eqn:D}).

It can  easily be shown that the 
additional terms appearing in the RHS of Eq.(\ref{eqn:inhomogeneous_wave}) 
are of order $M_t^2$,  where $M_t$ is the turbulent Mach number,
compared with the Reynolds source term (see also GK). 
As $M_t$ is small in the solar convection zone ($M_t
\lesssim 0.3$), these additional terms have been neglected.

The wave operator $\displaystyle { \left ( \partial  ^2 / \partial t ^2 - \vec
L \right )}$ acting on the turbulent velocity field $\vec u$  
in Eq. (\ref{eqn:inhomogeneous_wave}) gives rise to contributions 
which are  either negligible compared with 
 the Reynolds source term or of the same order. In this last case, 
the associated source term does not  contribute 
to the wave excitation  because it is  linear in terms of
the  turbulent fluctuations (see also Sect.~4.1). 
Therefore, in the LHS of Eq.(\ref{eqn:inhomogeneous_wave}), $\vec v$ can be replaced by $\vec v_{osc}$.

In deriving Eq.(\ref{eqn:vosc}), the time variation of the amplitude $A$ is neglected
since 
we assume, as in the solar case, that the eddy time correlation is of the 
order of a few minutes in the excitation region and 
the associated angular frequency is comparable to 
$\omega_0$ ($\sim 3 \, \mathrm{mHz} $), which is much larger 
than  the oscillation damping rate $\eta$ ($ \sim 100 \, \mu \mathrm{Hz})$. 

\subsection{Mean square amplitude}

Substituting Eq. (\ref{eqn:vosc}) into
 Eq. (\ref{eqn:inhomogeneous_wave}), gives, with the 
help of Eq. (\ref{eqn:homogeneous_wave}), 
\begin{eqnarray}
   2  \omega_0^2 \, \deriv{A}{t}  \,
\rho_0 \vec \xi  + 
i \omega_0 \, \vec {\cal D} \left (A \, \vec \xi  \right ) & & \nonumber \\  +
i \omega_0 \, \rho_0 \, \left ( \derivp  { ^2 } {t^2}  -  \vec L \right ) 
\left [ \frac{\rho_t}{\rho_0}  A \, \vec \xi    \right ] & = &  \derivp{\vec  {\cal S}}{t}
\label{eqn:dA_dt}
\end{eqnarray}
with
\eqna{
 \vec  {\cal S}  \equiv 
- \left ( \vec f_{t} + \vec \nabla  h_t \right )  	\; .
\label{eqn:St} 
}

As previously mentioned,  periods of  
oscillations are much shorter than their lifetime ($\sim
1/\eta$)  such that  $ \omega_0 \gg \eta$ and thus  $| d \ln A/dt | \ll
\omega_0$.  
Therefore the second derivative of $A$  arising in Eq. (\ref{eqn:dA_dt})
has been discarded.
Multiplying Eq. (\ref{eqn:dA_dt}) by $\rho_0 \vec \xi^*( \vec r , t )$ and  
integrating over the stellar volume gives
\begin{eqnarray}
  \deriv {A} {t} + \Delta \sigma~ A   = 
 \frac{1}{2 \omega_0^2 I} 
\int  d^3 x \,  \vec \xi^*  . \derivp{\vec  {\cal S} }{t}  \; , 
\label{eqn:dA_dt_2}
\end{eqnarray}
 where we have defined the complex quantity  
$\Delta \sigma = i\Delta \omega + \eta_D$ which comes from the contribution of
the operator ${\cal D} (\vec \xi)$. This expression is 
linear in the
eigenfunction $\xi$, hence $\Delta \omega$  corresponds 
to a  `dynamical' shift of the oscillation frequency due to turbulence. 
This shift is expected to be smaller or of the order of non-adiabatic 
frequency shifts, which are not considered here. The dynamical shift, $\Delta \omega$, will therefore be neglected. 
The real part, $\eta_D$,  
contributes to the dynamical damping. 
As in B92 and GK, $\eta_D$  is replaced in Eq. (\ref{eqn:dA_dt_2}) by the 
global damping rate $\eta$ in order to take into account all damping processes.
 The solution of Eq. (\ref{eqn:dA_dt_2}) 
 after integration by parts over the time is given by~: 
\eqna{
A(t)  =  \frac {i e^{-\eta t} } {2 \omega_0 I}  \int_{-\infty}^{t} dt^\prime   
\int d^3 x \,  e^{(\eta+i\omega_0) t^\prime} \,  
 \vec{\xi} (\vec x).  \vec {\cal S }  (\vec x, t^\prime) 
\label{eqn:A_t} 
}
and for the the  square amplitude: 
\eqna{
 |A|^2 (t) = \frac{e^{-2\eta t}} 
{4 (\omega_0 I)^2} \int_{-\infty}^{t} dt_1  dt_2  
\int d^3r_1  d^3r_2 
 e^{\sigma(t_1,t_2)} \nonumber \\  
\times \Bigl( \vec \xi (\vec{r}_1)\,  . \,
\vec{ {\cal S}
 (\vec{r}_1,t_1)\Bigr)  
\Bigl(\vec \xi (\vec{r}_2) \,  .  \, \vec {\cal S}^* (\vec{r}_2,t_2)\Bigr)
 }
\label{eqn:A2t}
}
with $\sigma(t_1,t_2) = \eta (t_1+t_2)+i\omega_0 (t_1-t_2)$.
Spatial integrations 
are  performed over the whole stellar envelope. 
For convenience we define the following coordinates
\eqn { \begin{array} {lcr}
\vspace{0.2cm} \vec x_0 =\displaystyle {  \frac {\vec r_2 + \vec r_1} {2}}  & & 
t_0 
= \displaystyle { \frac {t_1 +t_2} {2} } \nonumber \\
\vec r = \vec r_2 - \vec r_1 & & \tau = t_2 - t_1 \nonumber
\end{array}
}
$\vec x_0$  and $t_0$ are the average time-space position where the 
stochastic excitation is integrated whereas $\vec r $ and $\tau $ are
 related to the local turbulence.  $\tau$ corresponds to the fast time
 scale associated with the eddy lifetime and $t_0$ to the slow time
 scale associated with the oscillation growth rate ($\eta$).
In the following, $\vec{\nabla}_0 $ is the large-scale derivative associated 
with  $\vec{ x}_0 $,  $\vec{\nabla}_{\vec r} $ is the small-scale one associated with $\vec{r}$ and the derivative 
operators $\vec{\nabla}_1$ and $\vec{\nabla}_2 $ are associated 
with  $\vec r_1$ and $\vec r_2$ respectively.
The mean square amplitude  can be rewritten in terms 
of the new coordinates as
\eqna{
\left <  \mid A \mid ^2 (t) \right > &  = & \inv{4 (\omega_0 I)^2} 
\int_{-\infty}^{t} dt_0 e^{2 \eta (t_0-t)} 
\int_{2 (t_0-t)}^{2(t-t_0)} d\tau    \nonumber \\
& &\times \int  d^3x_0  d^3r \,  e^{-i\omega_0 \tau}  \nonumber \\
& & \hspace{-2cm} \times  \left < \,  \vec \xi \,   .  \,  \vec {\cal S}  [ \vec x_0-\frac{\vec r}{2}, t_0 - \frac{\tau} {2}] 
   \,   \vec \xi  \, .  \,  \vec {\cal S}^* [\vec x_0+\frac{\vec r}{2}, t_0 
+ \frac{\tau} {2}]  \,   \right > \; .
\label{eqn:A2t_2}
}
The operator $\langle .\rangle  $ denotes the statistical  average performed on 
an infinite number of independent realizations.

In the excitation region the eddy lifetime is much smaller than the
oscillation 
lifetime ($\sim 1/\eta$) such that the integration over $\tau$ can be
extended   to infinity. 
We assume  a stationary  turbulence, therefore the source 
term $\cal{S}$ is invariant over time $t_0$. 
 Integration over $t_0$ in Eq.(\ref{eqn:A2t_2}) and using the definition of $\vec {\cal{S}}$ in Eq.(\ref{eqn:St}) yields~:
\eqna{
\left < \mid A \mid ^2 \right >  = \frac{1}{8 \eta (\omega_0 I)^2}  
\left ( C_R^2 + C_S^2  \right )
\label{eqn:A2_0}
}
with 
\eqna{
C_R^2 & =& \int  d^3x_0   \,  \int_{-\infty}^{+\infty} d\tau \, e^{-i\omega_0 \tau}  \int d^3r  \nonumber \\  & & \times \left < \, 
\left (\rho_0 u_j u_i  \nabla_1^j  \xi^i \right )_1  
\left (\rho_0 u_j u_i \nabla_2^j  \xi^i \right )_2    
\right >  
\label{eqn:C2R}
}
the turbulent Reynolds stress contribution,
\eqna{
C_S^2 & =  &   \int  d^3x_0   \,  \int_{-\infty}^{+\infty} d\tau \, e^{-i\omega_0 \tau}  
\int d^3r   \nonumber \\ 
& &\times  \left < \, 
 \left ( h_t \, \vec \nabla \, .  \, \vec \xi      \right)_1  
\left (  h_t \,  \vec \nabla \,  . \, \vec \xi    \right )_2  \,
 \right >  
\label{eqn:C2S}
}
the entropy contribution.  
Subscripts   1 and 2 are the  values taken at the spatial and  temporal positions $[ \vec x_0-\frac{\vec r}{2}, - \frac{\tau} {2}]$  
and $ [\vec x_0+\frac{\vec r}{2}, \frac{\tau} {2}]$ respectively. 
We adopt the Einstein convention of summation upon repeated indices. 
Hereafter all time integrations over $\tau$ are understood to be performed 
in the range  $]-\infty , +\infty [$. 
 Note that the crossing term  between the Reynolds source term and the entropy source term does not contribute to the excitation 
(see Appendix~\ref{appendix:crossing_terms}).

\section{Reynolds stress contribution}

In the derivation of  Eq. (\ref{eqn:C2R}) and Eq. (\ref{eqn:C2S})  
integrations by parts have been performed in order for the gradient to 
act on the eigenfunction instead of
turbulent quantities.
We next suppose that the terms $(\rho_0 \nabla_1^i \xi^j ) $ 
and $(\rho_0 \nabla_2^i \xi^j)$ in Eq.(\ref{eqn:C2R})  
do not change  on the  length scale of the eddies. 
This implies  that $ \nabla_{\vec r}^i \xi^j \simeq 0 $.
Validity of this assumption  will be justified {\it a posteriori}
below.  
Consequently,  the Reynolds stress contribution can  be expressed as~: 
\eqna{
C_R^2 &  =  &\int d^3x_0  \,  \rho_0^2 \,  \nabla_0^i \xi^j \,  \nabla_0^l \xi^m  
 \nonumber \\
& & \times \int d\tau     d^3r \,  e^{-i\omega_0 \tau}    \langle u_i^\prime u_j^\prime u_l^\dprime u_m^\dprime \rangle
\label{eqn:A_Q2}
    } 
where  the following notations  have been used~:
\eqna{
\vec{u}^\prime = \vec{u}(\vec x_0-\frac{\vec r}{2},t_0-\frac{\tau}{2} ) & & 
\vec{u}^\dprime = \vec{u}(\vec x_0+\frac{\vec r}{2},t_0+\frac{\tau}{2} ) \; .
}
The  Quasi-Normal Approximation  
\cite[Chap VII-2, QNA hereafter]{Lesieur97} reduces the  
fourth-order velocity correlations as follows~:
\eqna{
  \langle u_i^\prime u_j^\prime u^\dprime_l u^\dprime_m \rangle  =  \langle u_i^\prime u_j^\prime  \rangle   \langle u^\dprime_l u^\dprime_m  \rangle + 
 \langle u_i^\prime u^\dprime_l \rangle    \langle u_j^\prime u^\dprime_m \rangle  \nonumber \\
+  \langle u_i^\prime  u^\dprime_m \rangle   \langle u_j^\prime u^\dprime_l \rangle \; .
\label{eqn:fourth-order}
}
This approximation  
 remains strictly valid for normally distributed fluctuating quantities.
{ As shown  by  \citet{Kraichnan57} and  \citet{Stein67} 
neglected terms can be large and therefore not negligible.  } 
Here the QNA  approximation
is  nevertheless assumed  valid as it is found justified 
when considering 3D simulations of the solar convection zone 
 \cite[work in progress]{Samadi00Phd}.

We denote  $\phi_{ij}(\vec k,\omega) $  as the  well known Fourier transform 
of the  second-order velocity correlations 
$\langle u_i^\prime u^\dprime_j \rangle $  \citep[e.g.][]{Stein67}.
For a stationary, incompressible, homogeneous and isotropic turbulence, the 
Fourier transform of the velocity correlation has the form \citep{Batchelor70} :
\eqna{
\phi_{ij}( \vec k,\omega) &  =& \frac { E( k,\omega) } { 4 \pi k^2}  \left( \delta_{ij}- \frac {k_i k_j} {k^2}  \right)
\label{eqn:phi_ij}
} 
where $ E( k,\omega) $ is the turbulent kinetic energy spectrum 
and $\delta_{ij}$ is the Kroenecker tensor.
 Following  \citet{Stein67}, the velocity
 energy spectrum $ E(k,\omega) $ is written as
\eqna{
E( k,\omega) =E(  k) \, \chi_k(\omega) \; .
\label{eqn:ek_chik}
}
 \citet{Stein67} and  \citet{Musielak94} suggest several forms for
the frequency factor. 
The  Gaussian function is the simplest choice and is defined as
\eqn{
\chi_k (\omega ) = \inv  { \omega_k \, \sqrt{\pi}}  e^{-(\omega / \omega_k)^2} \; .
\label{eqn:delta:omega}
}

The time spectrum is therefore 
normalized such that   \citep[Chap 8.1]{Tennekes82}
\eqn{
\int_{-\infty}^{+\infty} d\omega \,  \chi_k (\omega)  = 1 \; .
\label{eqn:chi_omega_norm}
}

One obtains from Eq.(\ref{eqn:fourth-order})~:
\begin{eqnarray}
 \int d\tau  \, d^3r  \,  e^{-i \omega_0 \tau}  \,   
\langle u_i^\prime u_j^\prime u_l^\dprime u_m^\dprime \rangle 
&  = & (2\pi)^4   \int d\omega \, d^3k \,  \nonumber \\
& & \hspace{-4.cm} \times  \left [ \phi_{il}(\vec k, \omega_0) \phi_{jm}(\vec k, \omega)  
+ \phi_{im}(\vec k, \omega_0) \phi_{jl}(\vec k, \omega) \right ] \; .
\label{eqn:int_phi_phi}
\end{eqnarray}

The first term in the 
RHS of Eq.(\ref{eqn:fourth-order}) when inserted into Eq. (\ref{eqn:int_phi_phi}) gives no contribution. 
With the help of Eq.(\ref{eqn:chi_omega_norm}) 
and Eq.(\ref{eqn:int_phi_phi}),
 the Reynolds stress contribution  can be written as 
\eqna{
C_R^2 & =  &\pi^{2}   \int  {d^3 x_0}  \,  
\left (\rho_0^2 \nabla_0^i \xi^j \nabla_0^l \xi^m \right )   
 \int d^3k \,  \nonumber \\ 
& & \times \left (  T_{ijlm} + T_{ijml} \right )  
\frac {E^2(k)} {k^4 }  \chi_k( \omega_0) 
\label{eqn:C2R_2}
 } 
where
$$
T_{ijlm} = \left( \delta_{il}- \frac {k_i k_l} {k^2}  \right)   \left( \delta_{jm}- \frac {k_j k_m} {k^2}  \right) \; .
$$
Excitation by convection takes place at the top 
of the convection zone. 
In this region, 
eigenvectors of acoustic modes with high radial order 
 can be considered essentially as radial and propagating in a plane-parallel manner.
Hence Eq.(\ref{eqn:C2R_2}) becomes 
 (see Appendix~\ref{appendix:anisoropy_effects})
\begin{eqnarray}
C_R^2 &=&  4 \pi^{3} \,  {\cal G}  \int_{0}^{M} dm   \,
\rho_0 \left (\deriv { \xi_r} {r} \right )^2  
\int_0^\infty dk \, \frac {E^2(k)} { k^2}  \chi_k ( \omega_0 )
\label{eqn:C2R_3}
\end{eqnarray}
where ${\cal G}$ is an anisotropic factor similar to 
Gough's \citeyearpar{Gough77}  one, as defined in  Eq. (\ref{eqn:phi}).  
Because of the term $\chi_k ( \omega_0 )$,  eddies with $\omega_k \gtrsim \omega_0 $ 
contribute the most to the integration over $k$, {\sl i.e.} 
to  mode excitation. 
This is in agreement with \citet{Goldreich90}, who state  that acoustic emission arises 
from eddies with $\omega_0 \tau_k \lesssim 1$.
 We have 
\eqn{
\frac {\omega_k} {\omega_0}  
\approx \frac{k u_k} {k_{osc} c_s }  \gtrsim \approx \frac{k}  {k_{osc} }  M_t \gtrsim 1 \; ,
\label{eqn:omega0_omegak}
}
where $k_{osc}$ is the wavenumber of the mode, $u_k$ the velocity of an eddy
with
 wavenumber $k$ and $M_t$ the turbulent Mach number, which is small. 
Eddies with $k \gg k_{osc}$ are those which contribute 
the most to modal excitation. 
The oscillation and the contributive eddies  are then well decoupled. 
Moreover, the stratification does not affect the
turbulent emission  \citep{Goldreich90}.
The above comments altogether justify  the 
assumption- at the beginning of this section-
  that $(\rho_0^2 \nabla_1^i \xi^j \nabla_2^k \xi^l)$
remains constant over the  length scale of the contributive eddies.
These conclusions also justify  the use of a homogeneous 
turbulence because the stratification and the oscillations have a 
characteristic scale length much larger than the contributive eddies.

\section{Contribution of entropy fluctuations}

As in Sect.~3 we use  $ \nabla_{\vec r}^i \xi^j \simeq 0 $  in
 Eq.(\ref{eqn:C2S}) to obtain~:
\eqna{
C_S^2 =   \int  d^3x_0  \, 
 \left ( \nabla_0^i \xi_i \right )^2 \int d\tau
 \, e^{- i \omega_0 \tau}  \int d^3r  \,
  \langle h_t ^\prime h_t ^\dprime \rangle     
\label{eqn:C2S_3}
}
where 
\eqn{
\begin{array}{lcr}
 h_t^\prime = h_t(\vec x_0-\frac{\vec r}{2},-\frac{\tau}{2} ) &  \textrm{and}&
 h_t^\dprime = h_t (\vec x_0+\frac{\vec r}{2},+\frac{\tau}{2} ) \; .
\end{array}
}
Integration over the time of the first term in the RHS of Eq. (\ref{eqn:h_t})  gives  
\eqn{
  h_t(\vec x ,t) =   -  \alpha_s  \delta s_t +  q_t(\vec x ,t)
\label{eqn:h_t_2}
}
with
\eqn{
q_t(\vec x ,t) \equiv
\int_{-\infty}^{t} dt^\prime \, \left  
( \vec \nabla. (\alpha_s \vec u \, s_t ) 
- s_t \vec u . \vec \nabla \alpha_s \right )  \; .
\label{eqn:h_t_3}
}
Contribution from the crossing term between 
the term $\alpha_s  \delta s_t$ (the linear term) and the term $q_t$ (the non linear  term), 
{\it i.e.} $ \langle \alpha_s  \delta s_t^\prime
q_t^\dprime \rangle$, 
vanishes (cf Appendix~\ref{appendix:crossing_terms}). 
We are left with
\eqna{
C_S^2 & = &  \int  d^3x_0  \, 
 \left ( \nabla_0^i \xi_i \right )^2 
\int d\tau \, e^{- i \omega_0 \tau}  
\int d^3r  \nonumber \\
& &  \times \Bigl[ \Bigl. \langle
\left ( \alpha_s  \delta s_t ^\prime \right ) \left ( \alpha_s  \delta s_t^\dprime \right ) \rangle     
+\langle
q_t^\prime q_t^\dprime
\rangle \Bigr] \Bigr  . \; .
\label{eqn:C2S_4}
}

\subsection{Contribution of the linear term}
\label{sec:linear_term}

We  consider the contribution of the Lagrangian entropy 
fluctuations ({\sl i.e.} the first
term in the RHS of Eq. (\ref{eqn:C2S_4})).

 In the Boussinesq approximation, the entropy fluctuation  $s_t$ is related to the temperature fluctuation $T_t$ as
\eqna{
s_t =  \left ( \derivp{s} {T} \right )_P  T_t \; .
\label{eqn:st_Tt}
}
In a free turbulent medium the Eulerian  temperature fluctuation acts as a 
passive scalar \cite[Chap VI-10]{Tennekes82,Lesieur97}. 
Thus one can expect that the  
Eulerian  entropy turbulent fluctuation acts as a passive scalar.

 Let $\phi_s(\vec k,\omega)$ be the Fourier transform of the  correlation product of the 
Eulerian entropy fluctuation. 
For any passive scalar, one has  the  relation  \cite[Chap V-10]{Lesieur97}
\eqna{
\phi_{s}(\vec k,\omega) = \frac {E_s (k,\omega) } {2 \pi k ^ 2} \; ,
\label{eqn:phi_s}
}
where the scalar spectrum $E_{s}( k,\omega) $ is  related 
to the scalar variance as  \cite[Chap V-10]{Lesieur97}
\eqna{
\frac{1}{2}  \langle s_t^2(\vec x_0,\omega) \rangle 
= \int_0^{\infty} dk \,  E_{s}(k,\omega)  \; .
}
 
Because the Lagrangian entropy
 fluctuation  acts as a passive scalar as
well, $\delta s_t$  scales as $s_t$.
The contribution of the Lagrangian entropy fluctuation 
is therefore proportional to 
\eqna{
\int_{-\infty}^{+\infty} d\tau \int d^3r  \,  e^{- i \omega_0 \tau} \, \langle s^\prime s^\dprime \rangle
 =  (2\pi)^4 \phi_s(\vec{0},\omega_0) \; ,
}
where the following notations have been defined~:
\eqn{
\begin{array}{ccc}
 s^\prime = s_t(\vec x_0-\frac{\vec r}{2},t_0-\frac{\tau}{2} ) & & 
 s^\dprime = s_t(\vec x_0+\frac{\vec r}{2},t_0+\frac{\tau}{2} ) \; .
\end{array}
}
As was done with the kinetic energy spectrum, 
the scalar energy spectrum $E_{s}( k,\omega)$ is decomposed such that 
\eqna{E_{s}( k,\omega) =E_{s}( k) \, \chi_k(\omega)
\label{eqn:esk_chik}
} 
with $\chi_k(\omega)$ the same frequency-dependent factor as in Sect.~3.
Let   $\chi_k(\omega)$ take the
Gaussian form~:  
as  $\omega_k \propto k u_k$ we have $ \chi_k(\omega_0)=0$ for $k= 0$. 
As $E_s(k)$ increases as $k^2$ in the vicinity of $k=0$ \cite[Chap
8.6]{Tennekes75,Tennekes82},
the quantity $ \phi_s(\vec{0},\omega_0)$ vanishes and so does  the linear 
term contribution. 
We emphasize that in the context of the approximation 
presented in Eq.(\ref{eqn:esk_chik}), no assumption has 
 been made regarding the behavior 
of the entropy energy spectrum $E_s(\vec k)$. Hence, 
this result remains  valid 
for any  scalar spectrum and for the velocity energy spectrum.

This result may also be explained as follows: in term of mode excitation,
 the linear entropy source term acts as a second-order correlation 
product $\langle s^\prime s^\dprime \rangle$.  Turbulence  and oscillation  
are coupled through the phase term  $e^{-i \omega_0 \tau}$ and through the 
turbulent time spectrum $\chi_k(\omega)$, which is the frequency-dependent 
component of $\langle s^\prime s^\dprime \rangle$ in the Fourier space. 
Therefore coupling between turbulence and  oscillation occurs at frequencies
 close to the oscillation frequency $\omega_0$ and thus involves 
eddies of wavenumber $k \gtrsim k_{osc} \, M_t^{-1}$ according to
 Eq. (\ref{eqn:omega0_omegak}). 
On the other hand, the spatial component of  $\langle s^\prime s^\dprime
 \rangle$ in the Fourier space favors eddies with the 
largest size ($k \rightarrow 0$).
These two opposite effects  clearly are incompatible and lead to vanishing of
 the entropy fluctuation contribution.

This does not happen for the contribution of the Reynolds  source term, 
which involves  the fourth-order velocity correlation product. 
According to the QNA this term  can be decomposed in terms of a product of two  
second-order velocity correlations.	
Coupling with the oscillation then becomes  non-linear and leads to
 an effective non-zero contribution. Thus, only non-linear 
terms can contribute to mode excitation while linear terms do not. 
This may be considered as a general result and justifies neglect of
several source  terms in section 2.2.

\subsection{Contribution of the non-linear terms} 

As the linear term does not contribute to the acoustic emission,
Eq.(\ref{eqn:C2S_4}) becomes, with Eq.(\ref{eqn:h_t_3}),
\begin{eqnarray}
C_S^2 &=&  \int  d^3x_0  \,  \left ( \vec \nabla_0 . \vec \xi  \right)^2 
    \,  \int d\tau   \,  e^{-i\omega_0 \tau}  \int d^3r  \nonumber \\
 & & \times \int_{-\infty}^{\tau/2} dt_3 \, \int_{-\infty}^{-\tau/2} \, dt_4 \, \nonumber \nonumber \\ 
 & & \times \left \{ \right.  \vec \nabla_1 \alpha_s . \, 
\left < \left (  s_t  \vec u  \right )_3   \,  
\left ( s_t  \vec u \right )_4  \right > . \vec \nabla_2 \alpha_s 
\\
&-& 2  \vec \nabla_1 \alpha_s . \left < \left ( s_t  \vec u \right )_3 \vec
 \nabla_2 . 
 \left ( \alpha_ s   s_t  \vec u  \right )_4  \right > 
\nonumber \\
&+& \left < \vec \nabla_1 . \left ( \alpha_s   s_t  \vec u  \right )_3  \,  \vec
 \nabla_2 .
  \left  ( \alpha_s  s_t  \vec u \right )_4 \right >  \left. \right  \} \nonumber
\label{eqn:C2S_5} 
\end{eqnarray}
where  subscripts $3$ and $4$ refer to evaluations at  positions $[\vec
x_0 - \vec r /2, t_3]$  and  $[\vec x_0 + \vec r /2,  t_4]$ respectively.
Similar way to the Reynolds contribution, we have  $\nabla_1^i \alpha_s \simeq 
 \nabla_2^i \alpha_s \simeq \nabla_0^i \alpha_s $. 
We use  an integration by parts and the
derivative operators $\nabla_1$ and $\nabla_2$ are replaced by the 
large-scale gradient $\nabla_0$. This finally gives 
\begin{eqnarray}
C_S^2 &=&  \int  d^3x_0  \,  \left ( \alpha_s \, \vec \nabla_0 . \vec \xi
\right)^2  \,
 g^{ij}   \,  \int d\tau \,   e^{-i\omega_0 \tau}  \int d^3r   \nonumber  \\
& & \times \int_{-\infty}^{\tau/2} dt_3 \,
\int_{-\infty}^{-\tau/2} \, 
dt_4 \,\left <  \left (  s_t u_i \right )_3 \, \left (  s_t  u_j  \right )_4  \right > 
\label{eqn:C2S_6} 
\end{eqnarray}
with
\eqna{
g^{ij} & =&  \nabla_0^i( \ln \mid \alpha_s \mid) \, \nabla_0^j( \ln \mid \alpha_s
\mid )  \nonumber \\
& &- 2 \nabla_0^i( \ln \mid \alpha_s \mid) \, \nabla_0^j(\ln \mid  \vec \nabla_0
. \vec \xi \mid )  \\
& & + \nabla_0^i(\ln \mid  \vec \nabla_0 . \vec \xi \mid ) \, 
\nabla_0^j(\ln \mid  \vec \nabla_0 . \vec \xi \mid ) \nonumber \; .
\label{eqn:gij}
}

Again, we consider the gradient of the stratification as radial 
and a plane parallel approximation in the excitation region. 
One obtains (see Appendix~\ref{appendix:entropy_contrib} and  Appendix~\ref{appendix:anisoropy_effects})
\begin{eqnarray}
C_S^2 &=&  \frac{ 4 \pi^3 \,{\cal H} } {\omega_0^2} \int d^3 x_0 
 \left ( \alpha_s \deriv{\xi_r}{r}   \right ) ^2   g_r    
    \, \nonumber \\ 
& & \times \int  dk \,  \frac { E_s(k)  E(k) } { k^2 } 
\int_{-\infty}^{+\infty} 
d\omega \, \chi_k(\omega_0+ \omega) \chi_k(\omega) 	
\label{eqn:C2S_7} 
\end{eqnarray}
with 
\eqna{
g_r =\inv{r^2} \left (  \deriv{} {\ln r}   \ln \mid \alpha_s \mid  - \deriv{} {\ln r}  \ln \mid  \deriv{} { r} \xi_r \mid  \right)^2
}
and  ${\cal H}$  an anisotropic factor similar to 
Gough's anisotropic factor (see Appendix~\ref{appendix:anisoropy_effects}).

The Mixing-Length Theory \citep[hereafter MLT][]{Bohm58,Cox68,Gough77} 
provides an estimate of the vertical velocity of the convective flow. 
The corresponding kinetic energy is  
transferred to smaller scales through the
turbulent cascade. The kinetic energy spectrum $E(k)$ is   normalized as
\eqn{  
\frac{1}{2} \, \langle u^2 (\vec x_0) \rangle  \equiv  
\int_0^{\infty}  dk \, E(k)  = \frac{1}{2} \,  \Phi w^2 = \frac{3} {2} \,  u_0^2  
\label{eqn:E:normalisation}
}
where $w$ is an estimate for the vertical convective velocity  
and  $\Phi$ is a factor  introduced by  \citet{Gough77} to take into account
anisotropy effects (Eq. \ref{eqn:phi}).  $u_0$ in the RHS of Eq.(\ref{eqn:E:normalisation}) has been introduced for convenience.

The MLT provides a relation between the  temperature fluctuations and 
the vertical convective velocity $w$  \citep{Gough77} :
\eqn{
w ^ 2 = \frac{g \Lambda \delta} {2 \Phi T} \,  \langle T_t^2\rangle^{1/2}
\label{eqn:w2_Theta}
}
where $\displaystyle{ \delta = \left ( \derivp{\ln \rho_0} {\ln T} \right )_P} $.
Thus Eq. (\ref{eqn:st_Tt}, \ref{eqn:w2_Theta}) 
provide an estimate of the entropy scalar variance
\eqn{
\tilde s^2 \equiv \langle s_t ^2 ( \vec x_0 ) \rangle 
 = \left ( \frac{2 \Phi C_P}{g \Lambda \delta } \right) ^2 w^4 \; .
\label{eqn:msqst_w4}
}
This enables us to normalize the entropy spectrum $E_s(k)$ 
\eqn{
\inv{2} \langle s_t ^2 ( \vec x_0 ) \rangle = \int_0^{\infty}  dk \, E_s(k) = \inv{2} \tilde s^2 \; .
\label{eqn:ES:normalisation}
}

Following GMK we define ${\cal R}$ as 
\eqn{
{\cal R} \equiv \frac{ \alpha_s   \tilde s  }  {\rho_0 u_0^2 } 
  = \frac{6}{\alpha_c}  \Gamma_1 \; ,
} where  $\alpha_c$ is the mixing-length parameter defined as usual 
 by $\Lambda = \alpha_c H_p$ with $\Lambda$ being the mixing-length and $H_p$ the
pressure scale height and $\Gamma_1$ is the usual adiabatic index.
The quantity ${\cal R}^2$  roughly  measures  the ratio of the excitation by 
 entropy fluctuations to that by fluctuations of  Reynolds stresses.

 The entropy contribution can then be written as~:
\eqna{
C_S^2 &  = & 4 \pi^{3}  \,{\cal H} \int_{0}^{M} dm  \, \rho_0 u_0^4  
\, {\cal R}^2 \, {\cal F}^2 \left (\deriv { \xi_r} {r} \right )^2  
\left ( \frac{u_0} {\Lambda \omega_0} \right )^2  \nonumber \\ 
& & \times \int
{dk \over k^2} \frac { E_s(k) E(k) } {\tilde s^2 u_0^2}  \int_{-\infty}^{+\infty}
 d\omega \,     
\chi_k ( \omega_0+ \omega ) \chi_k ( \omega )
\label{eqn:C2S_8}
}
with $\ds{{\cal F}^2  \equiv \Lambda^2 \,  g_r }$.

\section{Complete formulation}

With  Eq.(\ref{eqn:C2R_3}) and Eq.(\ref{eqn:C2S_8}),  the oscillation 
amplitude Eq.(\ref{eqn:A2_0}) is rewritten as
\eqna{
\left <\mid A \mid ^2 \right > = \frac{ \pi^{3}  }
{ 2  \eta (\omega_0 I)^2} \int_{0}^{M} dm \,  
\rho_0 u_0^4  \left (\deriv { \xi_r} {r} \right )^2 \nonumber 
\\ 
\times  \int_0^\infty {dk \over  k^2} \,  \frac {E(k)} {u_0^2}  F(k,\omega_0)
\label{eqn:A2}
}
where 
\eqna{
F(k,\omega_0) & \equiv&  {\cal G} \, \frac{E(k)}{u_0^2} \, \chi_k ( \omega_0 ) + 
 {\cal H}   {\cal R}^2 {\cal F}^2 \frac{E_s(k)}{\tilde s^2} \left ( \frac
{u_0}{\Lambda \omega_0}\right )^2
\nonumber \\ 
& & \times \int_{-\infty}^{+\infty} d\omega \, \chi_k ( \omega_0 + \omega)\chi_k (\omega
) \; ,
\label{eqn:A2_1}
}
where  ${\cal G}$ and ${\cal H}$ are given in Appendix~\ref{appendix:anisoropy_effects}.

Results in  \citet{Samadi00II}    for the solar case
 and in \citet{Samadi00III}  for Procyon
 are based on the above general expression for the mean 
square amplitude. In order to compare with results of previous
works, we next consider specific temporal and energy spectra, namely 
the Gaussian time spectrum and the Kolmogorov spectrum.

\subsection{Gaussian time spectrum}

Let consider the Gaussian time spectrum given by 
Eq.(\ref{eqn:delta:omega}). This time spectrum 
corresponds  in  the time space to   a gaussian function 
where  linewidth is  equal to  $2 \tau_k$ and $\tau_k$ is  
the characteristic time correlation length of an eddy  of  wavenumber $k$. 
Hence $\omega_k$ and $\tau_k$ are related to each other as
\eqn{
\omega_k= {2  \over  \tau_k} \; .
} 
As in B92 we define $\tau_k$  as 
\eqn{
\tau_k \equiv {\lambda  \over k u_k}
\label{eqn:tau_k}
} 
where the velocity  $u_k$ of the eddy with wave number $k$ 
is related to the kinetic  energy spectrum $E(k)$ as   \citep{Stein67} 
\eqn{
u_k^2 =  \int_k^{2 k}  dk \, E(k) \; .
\label{eqn:uk2}
} 
The factor $\lambda$ in Eq.(\ref{eqn:tau_k}) 
is introduced in order to gauge the definitions of  $\tau_k $ 
and  $u_k$  which involve some arbitrariness.

Let us define the wavenumber $k_0$ as the wavenumber of the largest eddy in 
the inertial range. Thus we relate $k_0$ to the mixing length as follows~:
\eqn{
k_0 =  \frac{2 \pi} {\beta \Lambda} 
\label{eqn:k_0}
}
where the parameter $\beta$ is introduced here again
in order to gauge 
the definition of $k_0$. 

For convenience we define the following   variables~:
\eqn{
\begin{array} {ccccc}
K =\displaystyle { \frac {k} {k_0} } &  , & u_K = \displaystyle { \frac { u_k} {u_0} } &  , &\omega_\Lambda = \displaystyle {  \frac{2 \pi u_0 } {\Lambda} } \; .
\end{array}
}
The mean square amplitude of Eq.(\ref{eqn:A2}) for the 
Gaussian form of  Eq. (\ref{eqn:delta:omega}) is given by 
\begin{eqnarray}
\left < \mid A\mid ^2 \right > &=&
 \frac{ \lambda \beta^4} { 32 \pi \eta (\omega_0 I)^2}
 \int_{0}^{M}  dm  \,  
\rho_0  u_0^3  \Lambda^{4}   \left ( \deriv{\xi_r}{r} \right )^2  
\nonumber \\ 
& &  \times  \left [{\mathcal G} \,  {\mathcal S}_R(X)  + {\mathcal H }
{ \mathcal R}^2 {\mathcal  F}^2  \,  {\mathcal S}_S(X) \right ] 
\label{eqn:A2_2}
\end{eqnarray}
with the ``source functions'' 
\begin{eqnarray}
{\mathcal S}_R(X) &=& 
\frac{1 } { 2 \sqrt{\pi} }   \int_0^\infty {dK \over K^3}  \, {1 \over u_K} 
\frac {k_0^2 \, E^2(K)} {u_0^4 } \,  e^{-Z^2} 
\label{eqn:SR} \\
{\mathcal S}_S (X) &=& \frac{X^{-2} } { 2  \left ( 2 \pi \right ) ^{5/2} }  
 \int_0^\infty {dK \over K^3}  {1 \over u_K} \nonumber \\ 
& & \times \Bigl(\frac {k_0 \, E_S(K) } { \tilde s^2 }\Bigr) \,
\Bigl(\frac {k_0 \, E(K)} {u_0^2}\Bigr)
 e^{-( Z \sqrt{2} )^2} \; ,
\label{eqn:SS} 
\end{eqnarray}
where $X(m,\omega_0) \equiv   \omega_0 / \omega_\Lambda $ and the ratio 
$Z = \lambda \beta  X /  2 K u_K$. As shown below, the ratios 
$k_0 \, E_S(K)/\tilde s^2 $ and $k_0 \, E(K)/u_0^2 $  only depend on $K$.

The  amplitude (Eq. \ref{eqn:A2_2}) is very sensitive to the
parameters $\beta$ and $\lambda$ because it  scales   
as $\lambda \beta^4$ and the quantity $\lambda \beta$ is 
involved in the exponential function of Eq.(\ref{eqn:SR}) and
Eq.(\ref{eqn:SS}). 
 In section \ref{sec:energy_spectrum} 
some physical arguments yield  a crude  estimate for the parameter $\beta$. 
Values of parameters $\lambda$, $\beta$ are discussed in some details in  \citet{Samadi00II}  in connection with solar seismic observations.

\subsection{Kolmogorov energy spectrum}
\label{sec:energy_spectrum}

The normalization condition
 Eq.(\ref{eqn:E:normalisation}) allows us 
to express the Kolmogorov spectrum  \cite[Chap VI-4]{Lesieur97}  as
\eqn{
E(K)=  \begin{array}{ccc}
\displaystyle { \frac{u_0^2} {k_0} }  \,  K^{-5/3}  & \mathrm{for} & K>1 \; .
\end{array}
\label{eqn:E:kolmogorov_2}
}

The expression for the mean square amplitude (Eq. \ref{eqn:A2}) involves 
also  the entropy  spectrum $E_s(k)$ which must therefore be determined. 
We recall that the energy spectra of the temperature and  of 
the entropy fluctuations  exhibit the same behavior 
(see section \ref{sec:linear_term}).
Turbulence theory predicts that the energy spectrum of the temperature
  fluctuations  
follows the same scaling law as that of the kinetic energy spectrum
  $E(k)$  
in the inertial-convective range  \cite[Chap VI-10]{Lesieur97}  
and it decreases as $k^{-17/3}$ in the inertial-conductive range  
because molecular-conductive effects are predominant. This behavior 
seems  to be quite well supported observations of solar granulation as is claimed 
by  \citet{Espagnet93} and \citet{Nesis93}.

We next turn to a specific case where 
we assume that the entropy energy 
spectrum lies in the inertial-convective range 
{\sl i.e.} the turbulent entropy spectrum scales as the kinetic spectrum.
According to the normalization conditions of Eq.(\ref{eqn:E:normalisation}) 
and of Eq.(\ref{eqn:ES:normalisation}), 
this assumption provides  the relation:
\eqn{
\frac { E_s(k,\omega) } {\tilde s^2}  =\frac{1} {3}  \frac { E(k,\omega) } {u_0^2} \; .
\label{eqn:E_s}
}

The source functions with Eq.(\ref{eqn:E:kolmogorov_2}) and   Eq.(\ref{eqn:E_s}), Eq. (\ref{eqn:SR}) 
and Eq. (\ref{eqn:SS}) become~:
\begin{eqnarray}
{\mathcal S}_R(X) &\simeq& 0.92 \int_{1}^{\infty} dK \, 
\frac {u_K^3} {K^5} \; e^{-Z^2} 
\label{eqn:SR_kolmo} \\
{\mathcal S}_S(X) &\simeq& \frac{0.008}{X^2} 
\int_{1}^{\infty} dK \, \frac {u_K^3} {K^5} \; 
e^{-( Z/\sqrt{2} )^2} 
\label{eqn:SS_kolmo}
\end{eqnarray}
with $X$ and $Z$ defined below Eq.(\ref{eqn:SS}).  
The source function $S_R(X)$ (Eq. \ref{eqn:SR_kolmo})
is similar to the source function $ S_Q(m,\omega)$ obtained  by
B92 (Eq. 2.20). 
However the source function for the entropy contribution $S_S(X)$
(Eq. \ref{eqn:SS_kolmo}) differs from the one established by B92 
(Eq. 3.9) in that the author extrapolated  
GK's formulation of the Reynolds stress contribution 
for the  contribution due to the Lagrangian entropy fluctuations.

For sake of comparison with the GMK formulation,
 we  simplify 
Eq.(\ref{eqn:SR_kolmo}) and Eq.(\ref{eqn:SS_kolmo})
by  using the fact that most of the 
stochastic emission occurs from eddies 
with $\omega_k \gtrsim \omega_0 $ (\ie with $e^{-(\omega_0/\omega_k)^2}  \sim 1$). 
For  $  \omega_0  / \omega_\Lambda \gtrsim 1 $ 
 integration   over $K$  leads then  to the asymptotic forms
\eqna{
\begin{array}{ccc}
\mathcal{S}_R \propto \, ( \omega_0 / \omega_\Lambda )^{-15/2}  & \mathrm{for} & \displaystyle{ \frac{\omega_0 } { \omega_\Lambda} \gtrsim 1 } \; ,
\end{array}
\label{eqn:SR_kolmo_approx}
}
\eqna{
\begin{array}{ccc}
\mathcal{S}_S \propto \, (\omega_0 / \omega_\Lambda )^{-19/2} &  \mathrm{for} & \displaystyle{  \frac{ \omega_0 } { \omega_\Lambda} \gtrsim 1 } \; .
\end{array}
\label{eqn:SS_kolmo_approx}
}
The asymptotic frequency dependence, Eq. (\ref{eqn:SR_kolmo_approx}),
 is as that found by GMK. On the other hand, 
 GMK assume the same frequency dependence for the two contributions
 while here  the source function $\mathcal{S}_S$ 
(Eq. \ref{eqn:SS_kolmo_approx})  corresponding to the entropy 
contribution exhibits a steeper slope. This result shows us that the power
 emission from the entropy contribution  is less efficient at high
 frequency compared to the Reynolds term and thus differs from the results of  GMK 
 but is  consistent with the GK statement that the contribution of the 
entropy fluctuation  is larger for the long period $p$-modes than for the 
short period $p$-modes.

\subsection{Constraints on free parameters}

We next turn to  the parameter $\beta$.  
The time scale at which the convective energy 
dissipates through  the  turbulent cascade is of order $\Lambda/w$.
Thus for  stationary turbulence and using Eq.(\ref{eqn:E:normalisation}), 
the rate of injection of kinetic energy $\epsilon$ \cite[Chap VI-3]{Lesieur97} can be crudely estimated as
\eqn{
\epsilon \simeq {3 \over 2} \, u_0^2 \, w / \Lambda \; .
\label{eqn:E:epsilon_2}
}
As $E(k) = C_K \, \epsilon ^{2/3} k^{-5/3}$ \citep[see][Chap VI-4]{Lesieur97} where $C_K$ is the Kolmogorov universal constant, which is close to $1.5$, and  from  Eq.(\ref{eqn:E:kolmogorov_2}) and Eq.(\ref{eqn:E:epsilon_2}) one obtains
\eqn{
k_0 \simeq  {3 \over 2} \, C_K^{3/2} (\Phi/3)^{-1/2} \Lambda^{-1}
}
and then
\eqn{
\beta \simeq {4\pi \over 3} \, C_K^{-3/2} (\Phi/3)^{1/2}  \; .
}
This suggests that crudely $\beta \simeq 1.9$ with $\Phi=2$,  
the value of the anisotropic factor consistent with BV's MLT.

The value suggested for $\beta$ is  somewhat approximate. 
Therefore, as in the case of the parameter $\lambda$, we consider $\beta$ as a free parameter.
However, the value of $\beta \lambda$ is constrained by an upper limit. Indeed, 
 let $\tau_{k_0}$ be the correlation time of the largest eddy in the inertial domain. The lifetime of the largest eddies in the inertial range cannot be longer than the characteristic time $\Lambda /w$  at which the convective energy dissipates into  the turbulent cascade.
Therefore we must have $\tau_{k_0} \lesssim \Lambda /w $ and according to Eqs.(\ref{eqn:tau_k},\ref{eqn:uk2}) evaluated for $k=k_0$ and Eqs.(\ref{eqn:k_0},\ref{eqn:E:normalisation}) we obtain   $\beta \lambda \lesssim 2.7 \,  \Phi^{1/2}  $.

\section{Discussion} 

In the present work, a formulation has been 
 established which yields the oscillation amplitude 
of a stellar oscillation mode when it is stochastically excited by turbulent
convection. 
 The main result of this paper is the expression for the mean square oscillation amplitude $\left <\mid A \mid ^2 \right > $ given by Eq.(\ref{eqn:A2}) and Eq.(\ref{eqn:A2_1}).
The derivation is based on theoretical developments of previous
 works (GK, GMK, B92) but an effort has been made 
to obtain a sufficiently general  expression  which enables one to
implement any type of turbulent (kinetic and entropic) spectra
and any type of  temporal spectra for the turbulent eddies.

For comparison purpose, we next focused on a gaussian temporal and Kolmogorov
energy spectra; 
  we then ended up with the same expression for the Reynolds stress contribution as
  obtained in GMK and B92.
We must stress however that in order to use 
 the same formulation   
for an energy spectrum other than the Kolmogorov one, such as for instance 
the Spiegel spectrum, a general expression such as 
 Eq. (\ref{eqn:A2}) must  be used.

As far as  the entropy contribution is concerned, 
we found that the 
linear term due to the entropy fluctuation gives no contribution
and  that it is  the advection of the Eulerian entropy fluctuations by 
the turbulent velocity field which  produces a nonzero acoustic emission.

In the derivation of the expected mean-square amplitude (Eq. \ref{eqn:A2}), 
several assumptions and approximations have been made. For instance, it has been assumed that the oscillations and the
 stratification are decoupled for the eddies which 
contribute to the stochastic power emission. This assumption was shown to 
be valid  and is in agreement with \citet{Goldreich90}. 
In addition, we have used the plane-parallel approximation, which is valid in 
the excitation region. 
Other assumptions are based on results from Stein's work 
\citeyearpar{Stein67}, 
such as the separation of the kinetic energy spectrum $E(k,\omega)$ in term 
of a purely spatial energy spectrum $E(k)$ and a time-dependent factor 
$\chi_k ( \omega )$ for an eddy of wavenumber $k$. 
As in \citet{Stein67} the QNA has been used.  
We have used crude approximations for estimations of the velocity 
and the life-time  of an eddy as proposed by   \citet{Stein67}. 
This led us to introduce  the  free  parameter $\lambda$ in
 the definition of  the  eddy lifetime.

The entropy fluctuation has been considered to act as a  passive 
scalar and we have extended the separation of the kinetic energy spectrum 
in terms of a purely spatial energy spectrum $E(k)$ and a time-dependent 
factor to the entropy energy spectrum. 

The MLT was required in order to estimate the power 
injected in the velocity and entropy turbulent cascade 
(Eq. \ref{eqn:A2_2}). This theory, which assumes the Boussinesq 
approximation, is well known to be a crude approximation. 
For instance the MLT predicts that  the characteristic size 
of the largest turbulent element  is comparable  with 
the scale height of the stratification. This is in contrast 
with the homogeneous hypothesis considered here for the  
description of the turbulent medium.	
However, it has already been stressed that the stochastic 
emission is not affected by the stratification. 
The use of an homogeneous turbulence is therefore valid.

The size of the largest eddy in the inertial range is estimated by the mixing length ($\Lambda$) 
according to the MLT. We have related the wavenumber of the largest 
eddy in the inertial range ($k_0$) to the mixing length.  
However, as for  the  eddy lifetime, this relation is rather arbitrary 
and therefore involves uncertainties.  We have  therefore introduced an additional  free 
 parameter $\beta$.

We have  considered an isotropic turbulence. However, effects of 
 anisotropy in the amplitude computaion have been partially taken into
 account. 
In this way, two anisotropic factors $\mathcal{H}$ and $\mathcal{G}$  
 have been introduced for both contributions respectively.  
These factors have been related to  \citet{Gough77} anisotropic factor $\Phi$.

It is possible to validate some of these  approximations by comparing them with results
of 3D simulations of the solar envelope \cite[work in progress]{Samadi00Phd}: for instance,
the QNA  is found to be reasonably valid.

To date, and in the solar case,  several possible turbulent spectra
 can be  investigated and compared with solar seismic observations
\citep{Samadi00II}. But we can anticipate 
that the entropy contribution will be dominant, as already  pointed out by GMK. 
Comparison with solar data allows us to calibrate the free parameters
 which in turn can be used to compute  oscillation power spectra for various 
solar-like oscillation stars.
Indeed, unlike the Sun, it is not possible to determine the 
turbulent spectra of  other stars from observations of 
the surface granulation. In the prospect of forthcoming  space seismic projects (COROT, MONS, MOST, EDDINGTON)
 comparison of theoretical computations with seismic data of several
 solar-like oscillation stars will provide useful 
constraints on  stellar turbulent spectra. 

\begin{acknowledgements}
We are indebted to A. Mangeney for particularly useful suggestions and advice without which this work would not have been possible.
We also thank  J.-P. Zahn, M. Rieutord and A. Baglin for useful discussions.
\end{acknowledgements}

\appendix
\section{Crossing terms} 	

\label{appendix:crossing_terms}

 The entropy  and  Reynolds source terms involve 
three  crossing terms arising from Eq. (\ref{eqn:A2t_2}). 
These  terms are  proportional to 
\eqna{
 \int d^3r \, \langle  s^\prime u_i^\dprime u_j^\dprime \rangle \\
 \int d^3r \,  \langle   u^\prime_l  s^\prime   u_i^\dprime u_j^\dprime  \rangle  \; .
}
The QNA  gives the following relation
\eqna{
\langle u_l^\prime s^\prime  u_i^\dprime u_j^\dprime \rangle = \langle u_l^\prime s^\prime \rangle \langle u_i^\dprime u_j^\dprime \rangle 
+   \langle u_l^\prime  u_i^\dprime \rangle  \langle s^\prime u_j^\dprime \rangle 
\nonumber \\ + \langle u_l^\prime  u_j^\dprime \rangle  \langle s^\prime u_i^\dprime \rangle \; .
\label{eqn:usuu}
}
For an isotropic, homogeneous and 
incompressible turbulence 
we have  $ \langle s^\prime u_j^\dprime \rangle =0$  
\cite[Chap V-8]{Lesieur97}. Thus the three terms 
in the RHS of Eq. (\ref{eqn:usuu}) vanish. 

The entropy source term introduces a crossing term 
between the linear and the non-linear terms of
 Eq.(\ref{eqn:h_t_3}). This term is  proportional to
\eqn{
 \int  d^3 r \,  \langle s^\prime \,  s^\dprime u_i^\dprime  \rangle \; .
}

Third  order correlation products of  
quantities following a normal distribution are expected to be equal
 to zero. Turbulent quantities, such as the  entropy fluctuations
  ($s_t$) and the turbulent velocity field ($\vec u$), are not 
  Gaussian random functions, since such turbulence would have
 no energy transfer between wavenumbers \citep{Lesieur97}. 
However, for a strongly turbulent medium, as in  a stellar  convection zone,  
approximation of  turbulent quantities by Gaussian random functions 
is consistent with the QNA  and 
can therefore be considered   as valid.

As a conclusion, the  crossing term $C_{RS}$  and 
the crossing terms between the linear and non-linear 
terms of the entropy contribution do not
 contribute to  oscillation excitation. 
\section{Entropy contribution } 	
\label{appendix:entropy_contrib}

The QNA provides the  relation
\eqna{
 \langle  u^i_3 u^j_4  s_3  s_4   \rangle
= \langle  u^i_3 u^j_4  \rangle  \langle  s_3 s_4   \rangle
+ \langle s_3  u^i_3  \rangle \langle s_4  u^j_4 \rangle 
\nonumber \\
+ \langle  s_3 u^j_4  \rangle \langle s_4 u^i_3  \rangle \; .
}
The terms of type $\langle u^i_3 s_3 \rangle $ and  $\langle u^i_3  s_4
\rangle$ 
 vanish  \cite[Chap V-8]{Lesieur97} and Eq. (\ref{eqn:C2S_6}) reduces to 
\begin{eqnarray}
C_S^2 &=&    \int d^3 x_0 \,    \left (\alpha_s  \vec \nabla_0 . \vec \xi
\right)^2    
 \, g_{ij} \, \int d\tau  \, e^{-i\omega_0 \tau} \int d^3r  
\nonumber  \\
 & & \times \int_{-\infty}^{\tau/2} dt_3 \int_{-\infty}^{-\tau/2} \, dt_4 \, 
\left <u_3^i u_4^j \right > \left < s_3 s_4 \right > .
\label{eqn:C2S_60} 
\end{eqnarray}

Stationarity of the turbulence allows us 
to write Eq.(\ref{eqn:C2S_60}) as
\begin{eqnarray}
C_S^2 & \propto &
 \int d\tau   \,  e^{-i\omega_0 \tau}  \, d^3r  \int_{-\infty}^{\tau/2} dt_3 
\nonumber \\
 & & \times  \int_{-\infty}^{-\tau/2} \, dt_4 \, 
\left <u^i u^j(t_4-t_3) \right > \left < s_t s_t (t_4-t_3) \right > 
\label{eqn:C2S_a1}
\end{eqnarray}
which can be expanded in the Fourier space as
\begin{eqnarray}
C_S^2  &\propto & (2 \pi) ^3  \int d\tau   \,  e^{-i\omega_0 \tau} \, d^3k
 \nonumber \\ 
& & \times \int_{-\infty}^{\tau/2} dt_3 \, \int_{-\infty}^{-\tau/2} \, dt_4 \, 
\int d\omega_3 \, d\omega_4    \nonumber \\ 
& &  \times \phi_{ij}(\vec k, \omega_3) \, \phi_s(\vec k, \omega_4)\, 
 e^{-i t_3 ( \omega_3+\omega_4)  }\,  e^{ i t_4 ( \omega_3+\omega_4)  }  .
\label{eqn:C2S_a2}
\end{eqnarray}
Integrating  over $t_3$ and $t_4$ in Eq.(\ref{eqn:C2S_a2}) and then over $\tau$ yields~:
\eqna{
C_S^2  \propto \frac{(2 \pi) ^4}{\omega_0^2}  \,  \int d^3k \, \int d\omega \, \phi_{ij}(\vec k, \omega) \, \phi_s(\vec k, \omega_0 + \omega) \; .
\label{eqn:C2S_61}
}
According to Eqs.(\ref{eqn:phi_ij}, \ref{eqn:ek_chik}, \ref{eqn:phi_s}, \ref{eqn:esk_chik}), Eq.(\ref{eqn:C2S_61}) can be written as
\eqna{
C_S^2  = \frac{(2 \pi) ^4}{\omega_0^2}  \, \int d^3 x_0 \,    \left (\alpha_s  \vec \nabla_0 . \vec \xi
\right)^2    
 \, g_{ij} \,    \int d\omega \,  \int d^3k \nonumber \\
\times \, \frac { E( k) } { 4 \pi k^2}  \left( \delta_{ij}- \frac {k_i k_j} {k^2}  \right) \, \frac {E_s (k) } {2 \pi k ^ 2} \,     
\chi_k ( \omega_0+ \omega ) \chi_k ( \omega ) \; .
\label{eqn:C2S_62}
}

\section{Anisotropy effects} 	
\label{appendix:anisoropy_effects}
We establish here the derivation of Eq. (\ref{eqn:C2R_3}) and
 Eq. (\ref{eqn:C2S_7})
 where the anisotropy factors $\mathcal G$ and $\mathcal H$ have been
 introduced.  Thus we evaluate these factors for several 
cases. Let $(x,y,z)$ be the real (anisotropic) coordinates
 and $(\tilde x, \tilde y, \tilde z)$ are isotropic coordinates. 
We  assume an isotropic turbulence in the horizontal layers 
while the anisotropy occurs in the vertical direction only.
The 
two sets of coordinates are then 
related to each other by the anisotropic factor $\mathcal{Q}$ as 
\eqn{
\begin{array} {ccccc} 
x = \mathcal{Q} \, \tilde x &  & y = \mathcal{Q} \, \tilde y & & z =   \tilde  z  \; .
\end{array}  
}
$\mathcal{Q}$ is also the ratio of the horizontal to vertical 
correlation lengths of the eddies. 
The equivalent relation in the wavenumber space is
\eqn{
\begin{array} {ccccc} 
k_x =  \mathcal{Q}^{-1} \, \tilde k_x  & & k_y = \mathcal{Q}^{-1} \,  \tilde k_y  & & k_z =  \tilde k_z 
\end{array}
}
where  $(k_x,k_y,k_z)$ and  $(\tilde k_x,\tilde k_y,\tilde k_z)$  are 
 anisotropic and isotropic wavenumber coordinates  respectively.
$\mathcal{Q}=1$ corresponds to an isotropic turbulence. 
We place ourselves in the case $\mathcal{Q} \neq 1$ but 
suppose that the energy spectra remain unchanged {\sl i.e.} 
consistent with an isotropic turbulence. 
We therefore  model only  geometrical effects due to  non-isotropic shapes of the eddies.

Evaluation of Eq. (\ref{eqn:C2R_2}) gives 
\begin{eqnarray}
C^2_R  &=&  4 \pi^3  \,  \int_{0}^{M} dm  \, \rho_0 \left (\deriv{\xi_r}{r}  
\right )^2  \nonumber \\  
& & \times  \int d k \,  \frac {E^2( k)} { k^2 } \,  \chi_k( \omega_0) \nonumber \\
& & \times  \int  d \theta \, \sin \theta  \left ( 1 - \frac{\mathcal{Q}^2 \cos^2
\theta }
 {(\mathcal{Q}^2-1) \cos^2 \theta + 1  } \right )^2 .
\end{eqnarray}

We define for convenience the factor $\mathcal{G}$ as 
\eqna{
\mathcal{G} =    \int_{-1}^1 d \cos \theta \, 
 \left ( 1 - \frac{\mathcal{Q}^2 \cos^2 \theta } {(\mathcal{Q}^2-1) 
\cos^2 \theta + 1  } \right )^2 \; .
\label{eqn:g}
}
This leads to  Eq. (\ref{eqn:C2R_3}).

 As was  done with the Reynolds stress contribution , it is  
straightforward
 to  obtain Eq. (\ref{eqn:C2S_7}) from Eq. (\ref{eqn:C2S_62}) 
where  the anisotropic factor $\mathcal{H}$ is 
\eqna{
\mathcal{H}   =  \int_{-1}^1 d \cos \theta \,  \left ( 1 - \frac{\mathcal{Q}^2 \cos^2 \theta } {(\mathcal{Q}^2-1) \cos^2 \theta + 1  } \right ) \; .
\label{eqn:h}
}
\citet{Gough77} defines the velocity anisotropic factor $\Phi$ as
\eqn{
\Phi = \frac{\langle u^2 \rangle }{\langle u_z^2 \rangle} \; .
\label{eqn:phi}
}
The lifetime of an eddy is  the same in any direction ; 
we then have
\eqna{
\frac{\mathcal{Q} \Lambda }
{  \langle u_x^2 \rangle^{1/2} } = 
\frac{\mathcal{Q} \Lambda }{  \langle u_y^2 \rangle^{1/2} } =
\frac{\Lambda }{  \langle u_z^2 \rangle^{1/2} } 
\label{eqn:correlation_time_any_direction}
}
where $\Lambda$  is the correlation length of the 
largest  eddy in the vertical direction.
Thus one obtains from
Eqs. (\ref{eqn:phi} , \ref{eqn:correlation_time_any_direction})  
a relation between $ \mathcal{Q}$ and $\Phi$
\eqna{
\Phi = 1 + 2 \, \mathcal{Q}^2 \; .
}
Integration of Eqs (\ref{eqn:g} , \ref{eqn:h}) gives
\eqna{
\mathcal{G}& = &  \frac{1} {  a^{5/2} } 
 \left [  \frac{2a^{3/2} - 3 \sqrt{a} } {a-1} - 3 \, \mathrm{atanh} ( \sqrt{a} )  \right ] \\
\mathcal{H} & = & \frac{ 2 }{ a^{3/2} } 
\left [\mathrm{atanh} ( \sqrt{a}) - \sqrt{a}  \right ] 
} with
\eqna{      
a= 1 - \mathcal{Q}^2 = \left ( 3 - \Phi \right ) /2 \; .
}
Table \ref{tab:gh}  gives values for
 $\mathcal{G}$ and $\mathcal{H}$ for different values of $\Phi$.

\begin{table}	
\caption{ Values for  $\mathcal{H}$ and $\mathcal{G}  $ for 
several cases:  the isotropic case corresponds to $\Phi=3$;
  $\Phi=2$ is consistent with  Bohm-Vitense's  MLT; $\Phi=5/3$ is 
the value that maximizes the convective heat fluxes  
\citep{Balmforth90a} and $\Phi=1.37$ is the value considered by \citet{Houdek01}.      
} 
\begin{center}
\begin{tabular}{lllll}
$\Phi$  & $\mathcal{Q}^2$ & $a$  & $\mathcal{G} $ & $\mathcal{H} $ \cr\hline
3 & 1 & 0 & 16/15 & 4/3 \cr
2 & 0.5 & 0.5 & 1.0 & 1.0 \cr
5/3 & 0.33 & 0.66 & 1.8 & 1.2 \cr
1.37 & 0.18 & 0.81 & 3.7 & 1.6 
\end{tabular}
\label{tab:gh}
\end{center}
\end{table}

\bibliography{../../biblio}

\begin{thebibliography}{33}
\expandafter\ifx\csname natexlab\endcsname\relax\def\natexlab#1{#1}\fi

\bibitem[{{Baglin} \& {The Corot Team}(1998)}]{Baglin98}
{Baglin}, A. \& {The Corot Team}. 1998, in IAU Symp. 185: New Eyes to See
  Inside the Sun and Stars, Vol. 185, 301

\bibitem[{{Balmforth}(1992)}]{Balmforth92c}
{Balmforth}, N.~J. 1992, \mnras, 255, 639

\bibitem[{{Balmforth} \& {Gough}(1990)}]{Balmforth90a}
{Balmforth}, N.~J. \& {Gough}, D.~O. 1990, \solphys, 128, 161

\bibitem[{Batchelor(1970)}]{Batchelor70}
Batchelor, G.~K. 1970, The theory of homogeneous turbulence (University Press)

\bibitem[{{B\"ohm - Vitense}(1958)}]{Bohm58}
{B\"ohm - Vitense}, E. 1958, Zeitschr. Astrophys., 46, 108

\bibitem[{{Christensen-Dalsgaard} \& {Frandsen}(1983)}]{JCD83a}
{Christensen-Dalsgaard}, J. \& {Frandsen}, S. 1983, \solphys, 82, 165

\bibitem[{{Cowling}(1941)}]{Cowling41}
{Cowling}, T.~G. 1941, \mnras, 101, 367

\bibitem[{Cox(1968)}]{Cox68}
Cox, J. 1968, Principles of stellar structure (Gordon and Breach)

\bibitem[{{Espagnet} {et~al.}(1993){Espagnet}, {Muller}, {Roudier}, \&
  {Mein}}]{Espagnet93}
{Espagnet}, O., {Muller}, R., {Roudier}, T., \& {Mein}, N. 1993, \aap, 271, 589

\bibitem[{{Favata} {et~al.}(2000){Favata}, {Roxburgh}, \&
  {Christensen-Dalsgaard}}]{Favata00}
{Favata}, F., {Roxburgh}, I., \& {Christensen-Dalsgaard}, J. 2000, in The Third
  MONS Workshop : Science Preparation and Target Selection, 49--54

\bibitem[{{Goldreich} \& {Keeley}(1977)}]{GK77}
{Goldreich}, P. \& {Keeley}, D.~A. 1977, \apj, 212, 243

\bibitem[{{Goldreich} \& {Kumar}(1990)}]{Goldreich90}
{Goldreich}, P. \& {Kumar}, P. 1990, \apj, 363, 694

\bibitem[{{Goldreich} {et~al.}(1994){Goldreich}, {Murray}, \& {Kumar}}]{GMK94}
{Goldreich}, P., {Murray}, N., \& {Kumar}, P. 1994, \apj, 424, 466

\bibitem[{{Gough}(1977)}]{Gough77}
{Gough}, D.~O. 1977, \apj, 214, 196

\bibitem[{{Houdek} {et~al.}(1999){Houdek}, {Balmforth},
  {Christensen-Dalsgaard}, \& {Gough}}]{Houdek99}
{Houdek}, G., {Balmforth}, N.~J., {Christensen-Dalsgaard}, J., \& {Gough},
  D.~O. 1999, \aap, 351, 582

\bibitem[{{Houdek} {et~al.}(2001){Houdek}, {Chaplin}, {Appourchaux},
  {Christensen-Dalsgaard}, {D\"appen}, {Elsworth}, {Isaak}, {News}, \&
  {Rabello-Soares}}]{Houdek01}
{Houdek}, G., {Chaplin}, W., {Appourchaux}, T., {Christensen-Dalsgaard}, J.,
  {D\"appen}, W., {Elsworth}, Y.~and{Gough}, D., {Isaak}, G., {News}, R., \&
  {Rabello-Soares}, M. 2001, submitted to MNRAS

\bibitem[{{Kjeldsen} \& {Bedding}(1998)}]{Kjeldsen98}
{Kjeldsen}, H. \& {Bedding}, T., eds. 1998, The First MONS Workshop: Science
  with a Small Space Telescope, ed. H.~{Kjeldsen} \& T.~{Bedding}, Aarhus
  Universitet.

\bibitem[{{Kraichnan}(1957)}]{Kraichnan57}
{Kraichnan}, R.~H. 1957, Physical Review., 107, 1485

\bibitem[{Lesieur(1997)}]{Lesieur97}
Lesieur, M. 1997, Turbulence in fluids (Kluwer Academic Publishers)

\bibitem[{{Lighthill}(1952)}]{Lighthill52}
{Lighthill}, M.~J. 1952, Proc. R. Soc. Lond., A211, 564

\bibitem[{{Matthews}(1998)}]{Matthews98}
{Matthews}, J.~M. 1998, in Structure and Dynamics of the Interior of the Sun
  and Sun-like Stars, 395

\bibitem[{{Musielak} {et~al.}(1994){Musielak}, {Rosner}, {Stein}, \&
  {Ulmschneider}}]{Musielak94}
{Musielak}, Z.~E., {Rosner}, R., {Stein}, R.~F., \& {Ulmschneider}, P. 1994,
  \apj, 423, 474

\bibitem[{{Nesis} {et~al.}(1993){Nesis}, {Hanslmeier}, {Hammer}, {Komm},
  {Mattig}, \& {Staiger}}]{Nesis93}
{Nesis}, A., {Hanslmeier}, A., {Hammer}, R., {Komm}, R., {Mattig}, W., \&
  {Staiger}, J. 1993, \aap, 279, 599

\bibitem[{{Osaki}(1990)}]{Osaki90}
{Osaki}, Y. 1990, in Lecture Notes in Physics : Progress of Seismology of the
  Sun and Stars, ed. Y.~{Osaki} \& H.~{Shibahashi} (Springer-Verlag), 75

\bibitem[{{Samadi}(2000)}]{Samadi00Phd}
{Samadi}, R. 2000, PhD thesis, Universit\'e Paris 6

\bibitem[{{Samadi} {et~al.}(2001{\natexlab{a}}){Samadi}, {Goupil}, \&
  {Lebreton}}]{Samadi00II}
{Samadi}, R., {Goupil}, M.-J., \& {Lebreton}, Y. 2001{\natexlab{a}}, \aap (in
  press)

\bibitem[{{Samadi} {et~al.}(2000){Samadi}, {Goupil}, {Lebreton}, \&
  {Baglin}}]{Samadi00b}
{Samadi}, R., {Goupil}, M.-J., {Lebreton}, Y., \& {Baglin}, A. 2000, in SOHO 10
  / GONG 2000 Workshop (in press)

\bibitem[{{Samadi} {et~al.}(2001{\natexlab{b}}){Samadi}, {Houdek}, {Goupil}, \&
  {Lebreton}}]{Samadi00III}
{Samadi}, R., {Houdek}, G., {Goupil}, M.-J., \& {Lebreton}, Y.
  2001{\natexlab{b}}, submitted to \aap

\bibitem[{{Stein} \& {Nordlund}(1991)}]{Stein91}
{Stein}, F. \& {Nordlund}, A. 1991, in Challenges to Theories of the Structure
  of Moderate-Mass Stars, Vol. 195 (ed D. Gough \& J. Toomre (Springer-Verlag))

\bibitem[{{Stein}(1967)}]{Stein67}
{Stein}, R.~F. 1967, Solar Physics, 2, 385

\bibitem[{{Tennekes}(1975)}]{Tennekes75}
{Tennekes}, H. 1975, Journal of Fluids Mechanics, 67, 561

\bibitem[{{Tennekes} \& {Lumley}(1982)}]{Tennekes82}
{Tennekes}, H. \& {Lumley}, J. 1982, A first course in turbulence, $8^{th}$
  edn. (MIT Press)

\bibitem[{{Unno} {et~al.}(1989){Unno}, {Osaki}, {Ando}, {Saio}, \&
  {Shibahashi}}]{Unno89}
{Unno}, W., {Osaki}, Y., {Ando}, H., {Saio}, H., \& {Shibahashi}, H. 1989,
  Nonradial oscillations of stars (Tokyo: University of Tokyo Press, 1989, 2nd
  ed.)

\end{thebibliography}
\bibliographystyle{apj}
\end{document}